\documentstyle[pra,aps,preprint]{revtex}

\begin{document}

\draft
\title{Intrinsic and operational observables in quantum mechanics}
\author{Berthold-Georg~Englert\cite{lmu} and Krzysztof
W\'odkiewicz\cite{uw}}
\address{Max-Planck-Institut f\"ur Quantenoptik,
Hans-Kopfermann-Strasse~1, D-85748~Garching, Germany}
\date{Received 23 June 1994}
\maketitle

\begin{abstract}%
The concept of intrinsic and operational observables in quantum mechanics
is introduced. In any realistic description of a quantum
measurement that includes a macroscopic detecting device, it is possible
to construct from the statistics of the recorded raw data a set of
operational quantities that correspond to the intrinsic quantum mechanical
observable. This general approach is illustrated by the example of an
operational measurement of the position and the momentum of a particle as
well as by an analysis of the operational detection of the phase of an
optical field. For the latter we identify the intrinsic phase operator
and report its explicit form.
\end{abstract}


\narrowtext

The quantum measurement theory provides for a conceptual framework
in which one can understand the features of the quantum world in
terms of measurable or observable quantities. Since the birth
of quantum physics, the theory of measurement has proved to
be controversial, both in its physical and philosophical
aspects. These controversies have generated long lasting debates
about the relation of the quantum formalism to the quantities that are
actually measured by macroscopic devices used in real experiments
\cite{wheeler}.

It is the purpose of this communication to present a general, down-to-earth
approach, connecting in a natural way the standard formalism of quantum
mechanics with the statistical raw data recorded in an experiment. In this
approach an operational link is established and discussed between the quantum
observables and the macroscopic devices used to detect and measure quantum
phenomena. We argue that, for each measurement, it is possible to construct
from the statistics of the recorded raw data  a set of operational quantities
that correspond to the quantum mechanical observables in a certain way.
Here, the ``raw data'' do not refer to the unprocessed laboratory records
but rather to the ``positive-operator-valued-measure'' or POVM that is the
mathematical representation of the statistical information gathered. In
one way of looking at quantum measurements~\cite{BLM} the emphasis is on
such POVMs. For us, however, the underlying intrinsic observable is the
heart of the matter.

We illustrate our approach to operational measurements using two different
examples. The first example deals with a model measurement of the position
and the momentum of a particle, and the second example is devoted to a
real homodyne detection of the phase of optical signals.

We start with a general description of our approach. For a quantum system
described by a density operator $\hat\rho$, statistical properties of an
arbitrary observable $\hat A$ can be evaluated with the aid of the
moment-generating function
\begin{equation}\label{defZ}
Z(\lambda)=\text{Tr}\{\hat\rho\exp(\lambda\hat A)\}
\end{equation}
in accordance with
\begin{equation}
\langle\hat A^n\rangle=\text{Tr}\{\hat\rho\hat A^n\}=
\frac{\text{d}^n}{\text{d}\lambda^n}Z(\lambda)\arrowvert_{\lambda=0}\,.
\end{equation}
Thus the generating function $Z(\lambda)$ contains all the relevant
statistical information about the system in state $\hat\rho$, but it makes no
reference to the apparatus employed in an actual measurement of the
observable $\hat A$ and its moments. To begin with, $Z(\lambda)$ is a purely
theoretical quantity; it is what would be measured in an ideal noise-free
measurement.

There are, however, numerous examples in the literature of measurements that
require realistic detecting devices. To name just a few, we mention the
quantum mechanical models of the ``pointer'' introduced by von Neumann
\cite{vneumann} and Arthurs and Kelly \cite{a+k}, their extension andrefinement
by Lamb \cite{Lamb}, the operational approach to the Heisenberg
microscope \cite{walls}, the quantum Zeno effect \cite{zeno}, the operational
phase-space in quantum mechanics \cite{kw84}, or the role of the apparatus
in the decoherence theory \cite{zurek}.

A realistic experiment necessarily involves additional degrees of freedom
which eventually enable the experimenter to convert the laboratory records
into a probability density, or rather a {\em propensity\/} density,
$\text{Pr}({a})$ of a classical variable $a$ \cite{propensity}. For this
purpose an analysis of the experimental setup is required, best perhaps in
the spirit of Lamb's operationalism \cite{Lamb}. The propensity thus found
determines classical averages as exemplified by
\begin{equation}
\overline{{a}^n}
=\int\!\text{d}{a}\,{a}^n\text{Pr}({a})\,.
\end{equation}
In the typical situations that we have in mind, the net effect of the
measuring device can be described by a ${a}$-dependent {\em filter\/}
$\cal F$, represented by a positive operator $\hat{\cal F}(a)$ such that
\begin{equation}
\text{Pr}({a})=k\,\text{Tr}\{\hat\rho\hat{\cal F}(a)\}
\end{equation}
where the coefficient $k$ is chosen in such a way that
$\int\!\text{d}{a}\,\text{Pr}({a})=1$.
In view of this linear relation, the requirement that
\begin{equation}\label{defOQO}
\overline{{a}^n}=\langle\hat A^{(n)}_{\cal F}\rangle
\end{equation}
holds for all $\hat\rho$, specifies a unique set of operators
$\hat{A}^{(n)}_{\cal F}$,
\begin{equation}
\hat A^{(n)}_{\cal F}=k\int\!\text{d}{a}\,{a}^n\hat{\cal F}(a)\,,
\end{equation}
for the given filter $\cal{F}$.

Inasmuch as the experimenter is guided by classical intuition when
designing the apparatus, we shall take for granted
that $\hat{A}=\hat{A}_{\cal F}^{(1)}$ holds and
that the quantum expectation value
$\langle\hat A^n\rangle$ agrees with the classical average
$\overline{{a}^n}$ in the correspondence limit. In other words, a good
measurement is characterized by the property that the classical
limits~\cite{clim} of $\hat A^n$ and $\hat A^{(n)}_{\cal{F}}$ are the same.

We shall employ the following terminology. We call $\hat A$ an intrinsic
quantum observable (IQO), whereas each $\hat A^{(n)}_{\cal{F}}$ is an
operational quantum observable (OQO). Thus, in the point of view that we wish
to advance, the measuring device $\cal{F}$ effectively replaces the powers of
IQOs by a set of OQOs. Rather than determining the generating function
$Z(\lambda)$ of Eq.~(\ref{defZ}), which refers to the IQO of interest, the
experimental results are compactly summarized in the filter-dependent
generating function
\begin{equation}
Z_{\cal{F}}(\lambda)=\sum_{n=0}^{\infty}
\frac{\lambda^n}{n!}\text{tr}\{\hat\rho\hat A^{(n)}_{\cal{F}}\}
=\int\!\text{d}{a}\,\exp({\lambda{a}})\text{Pr}({a})\,.
\end{equation}
The comparison with
\begin{equation}
Z(\lambda)=\int\!\text{d}A\,\langle A|\hat\rho|A\rangle \exp(\lambda A)
\end{equation}
shows that the probability distribution that is associated with the spectral
decomposition of $\hat A$ is effectively replaced by the propensity
$\text{d}{a}\,\text{Pr}({a})$, which refers to the filter $\cal{F}$ of the
measuring device.
Note that the quantity $k\,\text{d}{a}\,\hat{\cal F}(a)$ is the POVM
of the experiment in question. From our point of view, this POVM is not
interesting in itself; the filter function is merely necessary for the
identification of the OQOs, but the IQOs remain the objects of primary
interest.

There is then the obvious question: What is the relation between the OQOs and
the powers of the IQO? Two cases must be distinguished. First, we have the
standard situation in which the IQO is known, so that one just needs to
identify the OQOs corresponding to the filter of the actual measurement. The
noise introduced in the course of determining the propensity density
$\text{Pr}({a})$ can then be accounted for explicitly. In this way,
$Z(\lambda)$ can possibly be expressed in terms of $Z_{\cal F}(\lambda)$
whereafter the propensity has served its purpose. We shall illustrate this
standard case at a model of position and momentum measurements with respect
to a reference pointer in thermal equilibrium.

In the second case one deals with the unusual situation that the quantum
properties of the IQO are largely unknown, although the IQO has a well known
classical analog. The guidance provided by this classical analog suggests one
or more measurement schemes, each of which specifies a set of OQOs. While it
is clear that the looked-for IQO cannot be identified uniquely in such an
operational approach,
the choice $\hat{A}=\hat{A}_{\cal F}^{(1)}$ is certainly the most natural
one for the IQO associated with the OQOs of one experimental setup.
Once this IQO is identified, its $Z(\lambda)$ is available in principle and
can possibly be related to the generating function $Z_{\cal F}(\lambda)$ that
is determined experimentally. This second case is exemplified by the recent
measurements of the phase properties of optical fields by Noh, Foug\`eres,
and Mandel (NFM)~\cite{NFM}. Here the filter $\cal F$ accounts for
the beam splitters, mirrors, and photon counters used in the homodyne
detection. We shall treat this example  and identify the intrinsic phase
operator that corresponds most naturally to the OQOs defined by the NFM
apparatus.

As a rule, the algebraic properties of the $\hat A^{(n)}_{\cal F}$ operators
are quite different from those of the powers of $\hat A$. In particular, a
factorization is typically impossible, so that, for instance,
$\hat A^{(2)}_{\cal F}$ does not equal $(\hat A^{(1)}_{\cal F})^2$. The
operational spread
$\delta{a}=\bigl(\overline{{a}^2}-\overline{a}^2\bigr)^{1/2}=
\bigl(\langle\hat A^{(2)}_{\cal F}\rangle
-\langle\hat A^{(1)}_{\cal F}\rangle^2\bigr)^{1/2}$  is then different from
the quantum uncertainty $\Delta\hat A=\bigl(\langle\hat A^2\rangle
-\langle\hat A\rangle^2\bigr)^{1/2}$. Indeed, since the operational spread
$\delta{a}$ may refer to expectation values of two different operators,
its physical significance could be rather murky, in contrast to the quantum
uncertainty $\Delta\hat A$ with its familiar physical meaning. Further, it
is clear that the Heisenberg uncertainty relation obeyed by the product
$\Delta\hat A\,\Delta\hat B$ for two IQOs need not be equally valid for the
product $\delta{a}\,\delta{b}$ of the corresponding operational
spreads.

As an illustration of the general scheme we now turn to operational
measurements of the position and momentum of a particle in one dimension.
In particular, we consider a device that determines the overlap of the
density operator $\hat\rho$ of the system with the density operator of a
reference oscillator. This reference oscillator is supposed to be in a state
of thermal equilibrium with a temperature corresponding to $\bar n$
oscillator quanta. The oscillator also supplies natural units for distances
and momenta. Therefore, we shall take as the IQOs the dimensionless position
and momentum operators $\hat Q$ and $\hat P$ that refer to these oscillator
units. Now, in order to probe the system, the reference oscillator is
displaced both in position and in momentum by the amounts $q$ and
$p$, respectively. With these classical variables, the filter function
is
\begin{equation}
\hat{\cal F}({q},{p})=\exp(i{p}\hat{Q}-i{q}\hat{P})\,
\hat{\cal F}(0,0)\,\exp(i{q}\hat{P}-i{p}\hat{Q})\,,
\end{equation}
where
\begin{equation}
\hat{\cal F}(0,0)=\frac{1}{\bar{n}+1}
\left(\frac{\bar{n}}{\bar{n}+1}\right)^{(\hat{Q}^2+\hat{P}^2-1)/2}
\end{equation}
is the density operator of the reference oscillator when it is at rest and
located at the origin.

The propensity
$\text{Pr}({q},{p})=k\langle\hat{\cal F}({q},{p})\rangle$ is
normalized according to
$\int\!\text{d}{q}\,\text{d}{p}\,\text{Pr}({q},{p})=1$. The generating
function for the OQOs, for which
\begin{equation}
Z_{\cal F}(\lambda,\mu)=
\int\!\text{d}{q}\,\text{d}{p}\,
\exp(i\lambda{q}-i\mu{p})\text{Pr}({q},{p})
\end{equation}
is a convenient choice here, is then given by
\begin{eqnarray}
Z_{\cal F}(\lambda,\mu)&=&
\langle\exp(i\lambda\hat{Q}-i\mu\hat{P})\rangle\nonumber\\ &&\times
\exp\left(-\case{1}{4}(2\bar{n}+1)(\lambda^2+\mu^2)\right)\,.
\end{eqnarray}
The first factor can be regarded as a generating function $Z(\lambda,\mu)$
for expectation values of the intrinsic observables $\hat{Q}$ and $\hat{P}$,
and the second factor accounts for the noise that is unavoidably introduced
during the measurement.

With the generating function $Z_{\cal F}(\lambda,\mu)$ at hand we can proceed
to identify the operational observables. Upon expanding
$Z_{\cal F}(\lambda,\mu)$ in powers of $\lambda$ and $\mu$, the OQOs can be
read off in accordance with (\ref{defOQO}). For example, for those OQOs that
correspond to powers of $q$ only, this produces
\begin{equation}
\hat{Q}^{(n)}_{\cal F}=\left(\frac{1}{2i}\sqrt{2\bar{n}+1}\right)^n
H_n(i\hat{Q}/\sqrt{2\bar{n}+1})\,,
\end{equation}
where $H_n$ is the $n$-th Hermite polynomial. An analogous equation holds
for $\hat{P}^{(n)}_{\cal F}$. These relations can be
inverted in order to express the powers of $\hat{Q}$ and $\hat{P}$ in terms
of the OQOs whose expectation values are measured directly, as exemplified by
\begin{equation}
\hat{Q}=\hat{Q}^{(1)}_{\cal F}\,,\quad
\hat{Q}^2=\hat{Q}^{(2)}_{\cal F}
-(\bar{n}+\case{1}{2})\,,
\end{equation}
and so forth, and likewise for $\hat{P}^n$. An immediate consequence is the
analog of Heisenberg's uncertainty relation for the operational spreads,
viz.~\cite{kw87}
\begin{equation}\label{newHeis}
\delta{q}\,\delta{p}\geq\bar{n}+1\,,
\end{equation}
where the equal sign holds only if $\Delta\hat{Q}=\Delta\hat{P}=1/\sqrt{2}$.
Owing to the noise of the measuring device, the lower limit in
(\ref{newHeis}) is at least twice as large as that for the product of the
intrinsic uncertainties, $\Delta\hat{Q}\,\Delta\hat{P}\geq\case{1}{2}$.

As an illustration of the second case we now turn to the operational phase
difference of two  monochromatic electromagnetic waves determined by
measuring its sine and cosine simultaneously in a fittingly designed
interferometer. Such a device has been used in the recent NFM experiments
\cite{NFM} for a measurement of the quantum phase properties of a
low-intensity laser, relative to a high intensity classical field (local
oscillator). The experimental data are summarized in the so-called ``phase
distribution,'' which is nothing but the propensity density
$\text{Pr}(\varphi)$ for the classical phase variable $\varphi$ that NFM
associate operationally with the phase properties of the probe field.

By construction, this propensity is periodic,
$\text{Pr}(\varphi)=\text{Pr}(\varphi+2\pi)$, and we normalize it such that
\begin{equation}\label{normPr}
\int\limits_{(2\pi)}\text{d}\varphi\,\text{Pr}(\varphi)=1
\end{equation}
holds, where the integration covers any $\varphi$ interval of length $2\pi$.
The classical average of a periodic function $g(\varphi)=g(\varphi+2\pi)$ is
then given by
\begin{equation}\label{phiaver}
\overline{g(\varphi)}=
\int\limits_{(2\pi)}\text{d}\varphi\,
g(\varphi)\text{Pr}(\varphi)\,.
\end{equation}
This number equals the quantum expectation value
$\langle \hat{G}_{\cal F}\rangle$ of
the corresponding operational operator
$\hat{G}_{\cal F}(\hat{b}^{\dag},\hat{b})$, which is a
function of $\hat{b}^{\dag}$ and $\hat{b}$, the creation and annihilation
operators for photons in the probe field. It is obvious that any
$\hat{G}_{\cal F}$ of this kind is an OQO of the NFM experiment with the
filter ${\cal F}$ denoting the homodyne detection scheme used.

In the terminology of Ref.~\cite{JB+BGE91}, these OQOs are operators of the
phase --- {\em phasors\/}. In analogy to (\ref{defOQO}), the phasor basis
$\hat{E}_{\cal F}^{(n)}$ is thus identified by the defining property
\begin{equation}\label{NFMphasors}
\overline{\exp(in\varphi)}
=\left\langle\hat{E}_{\cal F}^{(n)}\right\rangle
\end{equation}
for $n=0,\pm1,\pm2,\ldots$~. The reality of the propensity density
$\text{Pr}(\varphi)$ implies that $\hat{E}_{\cal F}^{(-n)}$ is the adjoint of
$\hat{E}_{\cal F}^{(n)}$, and $\hat{E}_{\cal F}^{(0)}=1$ is an immediate
consequence of the normalization~(\ref{normPr}).

The members of the phasor basis are the basic OQOs because all other ones are
weighted sums of these fundamental OQOs. Indeed, a Fourier decomposition,
\begin{equation}\label{construct}
\hat{G}_{\cal F}=\sum_{n=-\infty}^{\infty}\hat{E}_{\cal F}^{(n)}
\int\limits_{(2\pi)}\frac{\text{d}\varphi}{2\pi}\,
\exp(-in\varphi)g(\varphi)\,,
\end{equation}
establishes the quantum counterpart $\hat{G}_{\cal F}$ to any periodic
function $g(\varphi)$.
This relation enables one to map  classical trigonometry onto the
corresponding quantum trigonometry associated with the NFM experiment. As an
example we have for the cosine and the (cosine)$^2$ functions these
operational definitions:
\begin{eqnarray}\label{cosines}
\hat{C}_{\cal F}^{(1)}&=&\case{1}{2}(\hat{E}_{\cal F}^{(1)}+
\hat{E}_{\cal F}^{(-1)})\,, \nonumber \\
\hat{C}_{\cal F}^{(2)}&=&\case{1}{4}(\hat{E}_{\cal F}^{(2)}+
2\hat{E}_{\cal F}^{(0)}+\hat{E}_{\cal F}^{(-2)})\,.
\end{eqnarray}
In fact, using relation (\ref{construct}) one can infer the entire quantum
trigonometry from the operational phasors. Note that, due to the operational
character of these cosine operators, they differ considerably  from the
Susskind-Glogower operators~\cite{sg64}, which are intrinsic in character.

The NFM experiment has been analyzed in two different, and largely
independent, ways. One analysis~\cite{MF+WS93,UL+HP93} found that the
propensity density $\text{Pr}(\varphi)$ is given by
\begin{equation}\label{phiPr}
\text{Pr}(\varphi)=\frac{1}{2\pi}\int\limits_0^{\infty}\!\text{d}I\,
\langle\beta|\hat{\rho}|\beta\rangle\,,
\end{equation}
where $\hat\rho$ is the density operator of the photon state of the probe
field and $|\beta\rangle$ is a normalized eigenstate of $\hat{b}$. Here,
$\beta=\sqrt{I}\exp(i\varphi)$ relates the eigenvalue $\beta$ to the
phase variable $\varphi$ and the intensity $I$. In the jargon of quantum
optics~\cite{WM}, $\text{Pr}(\varphi)$ is the radially integrated Q function
of $\hat\rho$, and $|\beta\rangle$ is a coherent state or Glauber state.

The other analysis \cite{PR+KW94} has identified the NFM phasors in normally
ordered form, compactly presented as~\cite{BGE+KW+PR}
\begin{equation}\label{phasorQ}
\hat{E}_{\cal F}^{(n)}=
\frac{(n/2)!}{n!}\,:M(n/2,n+1,-\hat{b}^{\dag}\hat{b}):\,\hat{b}^{n}
\end{equation}
for $n=0,1,2,\ldots$, where $M$ denotes the confluent hypergeometric
function, and the pair of colons indicates normal ordering of the operators
$\hat{b}^{\dag}$ and $\hat{b}$, that is: all $\hat{b}^{\dag}$'s to the left
of all $\hat{b}$'s. The connection between (\ref{phiPr}) and (\ref{phasorQ})
is implicitly contained in a 1974 paper by Paul~\cite{HP74}. A particularly
nice form of the basic phasors is~\cite{BGE+KW+PR,HP74}
\begin{equation}\label{phasorN}
\hat{E}_{\cal F}^{(n)}=
\frac{(\hat{b}^{\dag}\hat{b}+n/2)!}{(\hat{b}^{\dag}\hat{b}+n)!}\hat{b}^{n}
\quad\text{for}\quad n=0,1,2,\ldots\,;
\end{equation}
it is perhaps best suited for the construction of the OQOs associated with a
classical observable $g(\varphi)$.

The general procedure for finding the relations between the operationally
defined phasors and the intrinsic phase operator $\hat{\Phi}$ is not
applicable to the NFM experiment, simply because $\hat{\Phi}$ is unknown. It
can even be argued~\cite{JB+BGE91} that a unique phase operator does not
exist at all. There is a plethora of acceptable definitions which are all
equally good on general grounds. Nevertheless, the NFM experiment can be
analyzed, of course, and the  phasor basis (\ref{phasorN}) has been
identified as the OQOs.

 From this basis one can construct an operational phase operator
$\hat{\Phi}_{\cal F}$. We use relation (\ref{construct}) to  calculate the
weight factors of the phasors; these are just the  Fourier components of a
periodic function that is equal to the classical phase variable $\varphi$ in
an interval $\varphi_0<\varphi<\varphi_0+2\pi$~\cite{JB+BGE91}. The result
is:
\begin{eqnarray}
\hat{\Phi}_{\cal F}&=&
(\varphi_0+\pi)\hat{E}_{\cal F}^{(0)}\nonumber\\
&&+\sum_{n=1}^{\infty}\frac{i}{n}
\left(e^{-in\varphi_0}\hat{E}_{\cal F}^{(n)}
-e^{in\varphi_0}\hat{E}_{\cal F}^{(-n)}\right)\,.
\label{defPhi}
\end{eqnarray}
It is that  hermitean phase operator which is most naturally associated with
the NFM phase propensity, inasmuch as
\begin{equation}
\left\langle\hat{\Phi}_{\cal F}\right\rangle
=\int\limits_{\varphi_0}^{\varphi_0+2\pi}\text{d}\varphi
\,\varphi\text{Pr}(\varphi)
\end{equation}
equates the quantum expectation value of
$\hat{\Phi}_{\cal F}$ to the classical average of the
phase variable $\varphi$.
The specific choice made for the value of the constant $\varphi_0$ is without
physical significance, of course, so that the NFM experiment does not lead to
one single phase operator but rather to a family of closely related operators
labeled by the classical parameter $\varphi_0$. The spectrum of
$\hat{\Phi}_{\cal F}$ consists of all $\varphi$ values in said range. The
eigenstates of $\hat{\Phi}_{\cal F}$, however, are unknown as yet and
remain to be found.

One of the authors (K.~W.)  would like to thank Professor H.~Walther for his
invitation and for the hospitality extended to him at the
Max-Planck-Institut where this work has been done. This work was partially
supported by the Polish KBN Grant No.~20\,426\,91\,01.


\begin{references}
\bibitem[*]{lmu} Also at Sektion Physik, Universit\"at M\"unchen, Am
Cou\-lomb\-wall~1, D-85748~Garching, Germany.
\bibitem[\dag]{uw} Permanent address: Instytut Fizyki Teoretycznej,
Uniwersytet Warszawski, Ho\.za~69, PL-00-681~Warszawa, Poland.
\bibitem{wheeler} {\it Quantum Theory of Measurement\/}, edited by
J. H. Whee\-ler and W. H. Zurek (Princeton University Press, Princeton 1983).
\bibitem{BLM} See, for example, P. Busch, P. J. Lahti, and P. Mittelstaedt,
{\em The Quantum Theory of Measurement\/} (Springer-Verlag, Berlin 1991).
\bibitem{vneumann} J. von Neumann, {\em Mathematische Grundlagen der
Quantenmechanik\/} (Springer-Verlag, Berlin 1932).
\bibitem{a+k} E. Arthurs and J. Kelly, Bell System Tech.\ {\bf 44}, 725
(1965).
\bibitem{Lamb} W. E. Lamb Jr., Phys.\ Today {\bf 22}, 23 (1969); Ann.\ N. Y.
Acad.\ Sci.\ {\bf 480}, 407 (1986).
\bibitem{walls} P. Storey, M. Collett, and D. Walls, Phys.\ Rev.\ Lett.\
{\bf 68}, 472 (1992).
\bibitem{zeno} M. Itano, D. J. Heinzen, J. J. Bollinger, and
D. J. Wineland, Phys.\ Rev.\ A {\bf 41}, 2295 (1990); V. Frerichs and A.
Schenzle, Phys.\ Rev.\ A {\bf 44}, 1962 (1991).
\bibitem{kw84} K. W\'odkiewicz, Phys.\ Rev.\ Lett.\ {\bf 52}, 1064 (1984);
Phys.\ Lett.\ A {\bf 115}, 304 (1986).
\bibitem{zurek} W. H. Zurek, Physics Today {\bf 44}, 36 (October 1991).
\bibitem{propensity} About {\em propensity\/} or {\em potentiality\/} or
{\em tendency\/} see, for example K.~R.~Popper, {\em Quantum theory and the
schism in physics\/} (Hutchinson, London 1982), and W.~Heisenberg, {\em
Physics and Philosophy\/} (Harper\&Row, New York 1598).
\bibitem{clim} We use the term ``classical limit'' with the clear technical
meaning of Ref.~\cite{JB+BGE91}.
\bibitem{NFM} J. W. Noh, A. Foug\`eres, and L. Mandel, Phys.\ Rev.\ Lett.\
{\bf 67}, 1426 (1991); Phys.\ Rev.\ A {\bf 45}, 424 (1992); Phys.\ Rev.\
A {\bf 46}, 2840 (1992); Phys.\ Rev.\ Lett.\ {\bf 71}, 2579 (1993).
\bibitem{kw87} K. W\'odkiewicz, Phys.\ Lett.\ A {\bf 124}, 207 (1987).
\bibitem{JB+BGE91} J. Bergou and B.-G. Englert, Ann.\ Phys.\ (NY) {\bf 209},
479 (1991).
\bibitem{sg64} L. Susskind and J. Glogower, Physics {\bf 1}, 49 (1964).
\bibitem{MF+WS93} M. Freyberger and W. Schleich, Phys.\ Rev.\ A {\bf 47}, R30
(1993).
\bibitem{UL+HP93} U. Leonhardt and H. Paul, Phys.\ Rev.\ A {\bf 47}, R2460
(1993).
\bibitem{WM} See, for example, D. F. Walls and G. J. Milburn, {\em Quantum
Optics\/} (Springer-Verlag, Berlin 1994).
\bibitem{PR+KW94} P. Riegler and K. W\'odkiewicz, Phys.\ Rev.\ A {\bf 49},
1387 (1994).
\bibitem{BGE+KW+PR} B.-G. Englert, K. W\'odkiewicz, and P. Riegler, Phys.\
Rev.\ A (submitted).
\bibitem{HP74} H. Paul, Fortschr.\ Phys.\ {\bf 22}, 657 (1974).
\end{references}
\end{document}